\documentclass[preprint,aps,amsmath,showpacs,floatfix]{revtex4}

\begin{document}

\title{Controlling Phase Space Caustics in the Semiclassical Coherent State Propagator}

\author{A. D. Ribeiro$^\dag$$^\S$ and M. A. M. de Aguiar$^\S$}

\affiliation{
$^{\dag}$Instituto de F\'{\i}sica, Universidade de S\~ao Paulo,
CP 66318, 05315-970, S\~ao Paulo, SP, Brazil\\
$^{\S}$Instituto de F\'{\i}sica ``Gleb Wataghin'',
Universidade Estadual de Campinas, 13083-970, Campinas, SP, Brazil}

\begin{abstract}
The semiclassical formula for the quantum propagator in the coherent
state representation $\langle \mathbf{z}'' | e^{-i\hat{H}T/\hbar} |
\mathbf{z}'\rangle$ is not free from the problem of caustics. These
are singular points along the complex classical trajectories
specified by $\mathbf{z}'$, $\mathbf{z}''$ and $T$ where the usual
quadratic approximation fails, leading to divergences in the
semiclassical formula. In this paper we derive third order
approximations for this propagator that remain finite in the
vicinity of caustics. We use Maslov's method and the dual
representation proposed in Phys.~Rev.~Lett.~{\bf 95}, 050405 (2005)
to derive uniform, regular and transitional semiclassical
approximations for coherent state propagator in systems with two
degrees of freedom.

\end{abstract}

% 03.65.Sq Semiclassical theories and applications
% 31.15.Gy Semiclassical methods
% 02.30.Mv Approximations and expansions

\maketitle

%%%%%%%%%%%%%%%%%%%%%%%%%%%%%%%%%%%%%%%%%%%%%%%%%%%%%%%%%%%%%%%%%%%%
%%%%%%%%%%%%%%%%%%%%%%%%%%%%%%%%%%%%%%%%%%%%%%%%%%%%%%%%%%%%%%%%%%%%
\section{Introduction}

Semiclassical methods are the fundamental tool in the study of the
quantum-classical connection. In the limit where typical actions $S$
become much larger than Planck's constant $\hbar$, it is possible to
use classical ingredients, usually classical trajectories, to
produce approximations to quantum mechanical objects, like matrix
elements, wavefunctions and propagators. In Feynman's path integral
approach to quantum mechanics, semiclassical approximations consist
in realizing that the classical paths become dominant as $S >>
\hbar$ and it suffices to add together the contributions of a small
set of neighboring paths in the vicinity of the classical one. This
apparently simple procedure, however, has two well known caveats
that make the application of such formulas difficult: the existence
of non-contributing classical solutions and the presence of focal
points or caustics.

The first of these issues, which is not going to be further
discussed in this paper, is closely related to the {\em Stokes
Phenomenon}, which is the abrupt change in the number of
contributing solutions to an asymptotic formula when a certain
boundary in parameter space is crossed
\cite{stokes,phenomenon,bleistein}. Although a general criterion to
decide whether a trajectory should be included or not as a true
contribution to the formula exists, it is usually hard to verify in
practice. An example of a careful study of these solutions can be
found in \cite{parisio2}. More generally, one resorts to a simple
{\it a posteriori} criterion: the contribution of each trajectory is
computed and, if it leads to non-physical results, it is discarded.
This kind of prescription have been widely used in the last years
as, for example, in the semiclassical formula of the coherent state
propagator in one \cite{adachi} and two \cite{ribeiro1} spatial
dimensions, in the momentum propagator \cite{shudo} and in the
semiclassical evolution of gaussian wave packets \cite{Agu05}.

Singularities due to caustics is the other recurrent problem in
semiclassical formulas. In the WKB theory \cite{berrymount} the
semiclassical wave function in the position representation diverges
at the turning points $\dot q=0$. In the momentum representation the
equivalent problem occurs at the points where $\dot p=0$. In
addition, for the Van Vleck propagator, which is a semiclassical
formula of the propagator in the coordinate representation, $\langle
q'' | e^{-i\hat{H}T/\hbar} | q'\rangle$, singularities occur at the
focal points \cite{gutz}. These are points along the trajectory from
$q(0)=q'$ to $q(T)=q''$ where an initial set of trajectories issuing
from the same initial point $q(0)$ but with slightly different
momenta, get together again, focusing at the same point $q(t)$.

The failure of the semiclassical approximation at these points, and
the reason why a singularity develops there, is that the usual
quadratic approximation used to derive such formulas becomes
degenerate and third order contributions around the stationary
points become essential. The standard procedure to obtain improved
formulas valid at caustics is due to Maslov \cite{maslov} and it
consists of changing to a dual representation where the singularity
does not exist \cite{berry83,maslov}. For a singularity in
coordinates, one uses the momentum representation and vice-versa.
The trick is that, when transforming back to the representation
where the singularity exists, one should go beyond the quadratic
approximation, otherwise the singularity re-appears.

The subject of the present paper is the treatment of singularities
due to caustic in the semiclassical formula of the coherent state
propagator in two spatial dimensions
$\mathrm{K}(\mathbf{z}''^*,\mathbf{z}',T)\equiv\langle \mathbf{z}''
| e^{-i\hat{H}T/\hbar} | \mathbf{z}'\rangle$. In spite of the fact
that this is a phase space representation, where no turning points
exist, this propagator is not free from caustics
\cite{adachi,Klau95,tanaka98,ribeiro1}, although earlier works on
the subject indicated so \cite{mcdonald,klauder1,leboeuf,voros}.
These points have been termed {\em phase space caustics}.

The caustics in $\mathrm{K_{sc}} (\mathbf{z}''^*, \mathbf{z}',T)$
have the same origin as the focal point divergence in the Van-Vleck
propagator, namely, the breakdown of the quadratic approximation.
Therefore, it is natural to seek for a dual representation as in
Maslov's method to derive higher order approximations. However,
since both coordinates and momenta are used in the coherent states,
there seems to be no room for a natural dual representation. In a
recent paper \cite{prl} we have proposed the construction of an
application between $f(z^*)=\langle z|\psi \rangle$ and an associate
function $\tilde{f}(w)$ that plays the role of the dual
representation for the coherent state propagator. Using this
auxiliary mapping we were able to derived a uniform approximation
for the propagator of one-dimensional systems that is finite at
phase space caustics. In this paper, we use such a representation to
derive regular, transitional and uniform semiclassical approximation
for the coherent state propagator of two-dimensional systems, which
is the simplest case where conservative chaos is possible. The
resulting formulas involve, as expected, the Airy function and the
third derivatives of the action function.

This article is organized as follows: in Sect.~II we review the
semiclassical formula for the coherent state propagator in two
dimensions and discuss its singularities. In Sect.~III we review the
dual representation proposed in Ref.~\cite{prl} and generalize it
for two-dimensional systems. In Sect.~IV we use this representation
and the Maslov method to derive regular, transitional and uniform
approximations for the coherent state propagator. Our conclusions
and final remarks are presented in Sect.~V.

%%%%%%%%%%%%%%%%%%%%%%%%%%%%%%%%%%%%%%%%%%%%%%%%%%%%%%%%%%%%%%%%%%%%
%%%%%%%%%%%%%%%%%%%%%%%%%%%%%%%%%%%%%%%%%%%%%%%%%%%%%%%%%%%%%%%%%%%%
\section{The Semiclassical Limit of the Coherent State Propagator}

In this section we briefly discuss the usual semiclassical formula for the
propagator in the coherent state representation. The 2-D {\em
non-normalized} coherent state $|\mathbf{z}\rangle$ is the direct
product of two 1-D states, $|\mathbf{z} \rangle \equiv |z_x \rangle
\otimes |z_y \rangle$, where
\begin{equation}
\begin{array}{l}
\displaystyle{|z_r\rangle =
   e^{z_r \hat{a}_r^{\dagger}}
   |0\rangle ,}\\ \\
\displaystyle{\hat{a}_r^{\dagger} =
   \frac{1}{\sqrt{2}}
   \left(
     \frac{\hat{q}_r}{b_r}
     - i
     \frac{\hat{p}_r}{c_r}
   \right),}\\ \\
\displaystyle{z_r =
   \frac{1}{\sqrt{2}}
   \left(
     \frac{\bar{q}_r}{b_r}
     + i
     \frac{\bar{p}_r}{c_r}
   \right).}
\end{array}
\end{equation}
The index $r$ assumes the values $x$ or $y$. $| 0 \rangle$ is the
ground state of a harmonic oscillator of frequency $\omega_r=\hbar/(m
b_r^2)$, $\hat{a}_r^{\dagger}$ is the creation operator and
$\bar{q}_r $, $\bar{p}_r $ are the mean values of the position
$\hat{q}_r $ and momentum $\hat{p}_r $ operators, respectively. The
widths in position $b_r$ and momentum $c_r$ satisfy $b_r c_r =
\hbar$. In addition, the complex number $z_r$ is eigenvalue of
$\hat{a}_r$ with eigenvector $|z_r\rangle$.

The coherent state propagator $\mathrm{K}({\mathbf{z}''^*},
\mathbf{z}', T) \equiv \langle \mathbf{z}'' | e^{-i\hat{H}T/\hbar} |
\mathbf{z}' \rangle$ represents the probability amplitude that the
initial coherent state $|\mathbf{z}' \rangle$ evolves into another
coherent state $|\mathbf{z}''\rangle$ after a time $T$, according to
the Hamiltonian $\hat{H}$. Notice that, since the initial and final
coherent states are non-normalized, all the propagators considered
in this paper should be multiplied by $e^{-\frac{1}{2}|\mathbf
z'|-\frac{1}{2}|\mathbf z''|^2}$ to get the usual propagators with
normalized bras and kets.

The semiclassical approximation for this propagator was firstly
considered by Klauder \cite{Klau78,Klau79,Klau87a} and Weissman
\cite{Weis82b}. More recently, however, a detailed derivation
was presented for systems with one degree
of freedom \cite{Bar01}. The expression for two-dimensional systems
is \cite{ribeiro1}
\begin{equation}
\label{eq1} \mathrm{K}_{\mathrm{SC}}^{\mathrm{(2)}} \left(
\mathbf{z}''^*, \mathbf{z}', T  \right) =
   \sum_{\mathrm{traj.}}
   \sqrt{ \frac{1}{\left| \det \mathbf{\mathrm{M}_{vv}} \right|}} \,
   \exp{ \left\{ \frac{i}{\hbar} \, \mathcal F \right\}},
\end{equation}
where the index $^{\mathrm{(2)}}$ was inserted to indicate
explicitly that this formula was obtained by means of a second order
saddle point approximation. The factors $\mathbf{\mathrm{M}_{vv}}$
and $\mathcal F$ depend on (generally complex) classical trajectories. These
trajectories are best represented in terms of new variables
$\mathbf{u}$ and $\mathbf{v}$, instead of the canonical variables
$\mathbf{q}$ and $\mathbf{p}$, defined by
\begin{equation}
\label{eq2} u_r =
    \frac{1}{\sqrt{2}}
    \left( \frac{q_r}{b_r} + i \frac{p_r}{c_r} \right)
    \qquad \mathrm{and} \qquad
v_r =
    \frac{1}{\sqrt{2}}
    \left( \frac{q_r}{b_r} - i \frac{p_r}{c_r} \right) .
\end{equation}
The sum in Eq.~(\ref{eq1}) runs over all trajectories governed by
the complex Hamiltonian $\tilde{H} (\mathbf{u}, \mathbf{v}) \equiv
\langle \mathbf{v} | \hat{H} | \mathbf{u} \rangle$. They must
satisfy the boundary conditions $\mathbf{u}(0) \equiv \mathbf{u}' =
\mathbf{z}'$ and $\mathbf{v}(T) \equiv \mathbf{v}'' =
\mathbf{z}''^*$. Notice that $q_r$ and $p_r$ are complex variables,
while the propagator labels ($\bar{q}'_r$, $\bar{p}'_r$ for the
initial state and $\bar{q}''_r$, $\bar{p}''_r$ for the final one)
are real. In Eq. (\ref{eq1}), $\mathcal F$ is given by
\begin{equation}
\mathcal F( \mathbf{v}'',   \mathbf{u}', T) =
   \mathcal{S}( \mathbf{v}'',   \mathbf{u}', T) +
   \mathcal{G}( \mathbf{v}'',   \mathbf{u}', T) -
   \frac{\hbar}{2} \sigma_{\mathbf{vv}} ,
\label{eq3}
\end{equation}
where $\mathcal{S}$, the complex action of the trajectory, and
$\mathcal{G}$ are
\begin{eqnarray}
\mathcal{S}( \mathbf{v}'',   \mathbf{u}', T) &=&
   \int_{0}^{T}
   \left[
      \frac{i \hbar}{2}
      \left( \mathbf{ \dot{u} \, v - u \, \dot{v}} \right) -
      \tilde{H}
   \right] dt
   - \frac{i \hbar}{2} \left[ \mathbf{  u'' v'' + u'v' } \right],
\label{action}\\
\mathcal{G}( \mathbf{v}'',  \mathbf{u}', T) &=&
   \frac{1}{2} \int_{0}^{T}
   \left(
      \frac{\partial^{2}\tilde{H}}{\partial u_{x} \; \partial v_{x}}
      +
      \frac{\partial^{2}\tilde{H}}{\partial u_{y} \; \partial v_{y}}
   \right) \, dt \; .
\label{eq4}
\end{eqnarray}

\noindent The matrix $\mathbf{\mathrm{M}_{vv}}$ is a block of the
tangent matrix defined by
\begin{eqnarray}
\left(
   \begin{array}{c}
   \delta \mathbf{u}''\\
   \delta \mathbf{v}''\\
   \end{array}
\right) = \left(
   \begin{array}{cc}
   \mathbf{\mathrm{M}_{uu}}   & \mathbf{\mathrm{M}_{uv}}   \\
   \mathbf{\mathrm{M}_{vu}}   & \mathbf{\mathrm{M}_{vv}}   \\
   \end{array}
\right) \left(
   \begin{array}{c}
   \delta \mathbf{u}'\\
   \delta \mathbf{v}'\\
   \end{array}
\right) \, , \label{eq5}
\end{eqnarray}
where $\delta \mathbf{u}$ and $\delta \mathbf{v}$ are small
displacements around the complex trajectory. We use a single
(double) prime to indicate initial time $t=0$ (final time $t=T$).
The elements of the tangent matrix can be written in terms of second
derivatives of the action (see Ref. \cite{ribeiro1}). Finally,
$\sigma_{\mathbf{vv}}$ is the phase of $\det \mathbf{\mathrm{M}_{vv}}$.

The classical trajectories contributing to the propagator are
functions of nine real parameters: four initial labels $\bar{q}'_x$,
$\bar{q}'_y$, $\bar{p}'_x$ and $\bar{p}'_y$, four final labels
$\bar{q}_x''$, $\bar{q}''_y$, $\bar{p}''_x$ and $\bar{p}''_y$, and
the propagation time $T$. As one changes one of these parameters
continuously, it might happen that two independent solutions become
very similar to each other. In the limit situation they might
coalesce into a single trajectory, characterizing a phase space
caustic, or a bifurcation point. At the immediate neighborhood of
the caustic these solutions will satisfy identical boundary
conditions. Therefore, close to the caustic, we can set small
initial displacements $\delta \mathbf{u}' = 0 $ and $\delta
\mathbf{v}' \neq 0$ in such a manner that, after the time $T$,
$\delta \mathbf{u}'' \neq 0 $ and $\delta \mathbf{v}'' = 0$.
Eq.~(\ref{eq5}) then reduces to
\begin{equation}
\left(
   \begin{array}{c}
   \delta \mathbf{u}''\\
   0 \\
   \end{array}
\right) = \left(
   \begin{array}{cc}
   \mathbf{\mathrm{M}_{uu}}   & \mathbf{\mathrm{M}_{uv}}   \\
   \mathbf{\mathrm{M}_{vu}}   & \mathbf{\mathrm{M}_{vv}}   \\
   \end{array}
\right) \left(
   \begin{array}{c}
   0 \\
   \delta \mathbf{v}'\\
   \end{array}
\right) \, , \label{eq6}
\end{equation}
implying that $\det \mathbf{\mathrm{M}_{vv}} =0$. Therefore, at the
caustic the pre-factor of the Eq.~(\ref{eq1}), $|\det
\mathbf{\mathrm{M}_{vv}}|^{-1/2}$, diverges and the semiclassical
formula cannot be used. The main purpose of this paper is to correct
the semiclassical formula in these situations, replacing the
singular pre-factor by a well behaved Airy-type function.

As mentioned in the Introduction, in addition to the divergence of
the pre-factor, the semiclassical formula (\ref{eq1}) can exhibit
other peculiarities, which we shall not address here. For example,
for some complex trajectories the imaginary part of $\mathcal F$ can
be large and negative, giving unreasonably large contributions to
the propagator. This problem is related to the Stokes Phenomenon,
and lead to the exclusion of some trajectories from the sum in
Eq.~(\ref{eq1}) in order to eliminate the unphysical results they
produce \cite{adachi,Klau95,shudo,Agu05,ribeiro1,parisio2}.

%%%%%%%%%%%%%%%%%%%%%%%%%%%%%%%%%%%%%%%%%%%%%%%%%%%%%%%%%%%%%%%%%%%%
%%%%%%%%%%%%%%%%%%%%%%%%%%%%%%%%%%%%%%%%%%%%%%%%%%%%%%%%%%%%%%%%%%%%
\section{Dual Representation for the Coherent State Propagator}

The main difficulty in dealing with phase space caustics is the lack
of a dual representation for the coherent state propagator. Caustics
in position representation can be removed by changing to the
momentum representation and vice-versa. For the coherent state
propagator, since both position and momentum are being used, it is
not clear what to do to bypass the phase space caustics. In
Ref.~\cite{prl} we defined an application that plays the role of a
dual representation for the coherent state basis and used it to find
a uniform formula for the semiclassical propagator for
one-dimensional systems. The purpose of this section is to extend
the formalism of Ref.~\cite{prl} for systems with two degrees of
freedom.

Based on the relations
\begin{equation}
\mathbf{u}(T) \equiv \mathbf{u}'' =
   \frac{i}{\hbar}
   \frac{\partial \mathcal S}{\partial \mathbf{v}''}
\quad \mathrm{and} \quad \mathbf{v}(0) \equiv \mathbf{v}' =
   \frac{i}{\hbar}
   \frac{\partial \mathcal S}{\partial \mathbf{u}'} \, ,
\label{ull}
\end{equation}
which can be demonstrated by differentiating the complex action of
Eq.~(\ref{action}), we perform a Legendre transform on $\mathcal{S}
\left( { \mathbf{v}''}, \,  \mathbf{u}',T \right)$ replacing the
independent variable $\mathbf{v}''$ by $\mathbf{u}''= (i/
\hbar)(\partial \mathcal S /\partial \mathbf{v}'')$. The transformed
function $\tilde{\mathcal{S}}$ depends on the variables
$\mathbf{u}'$ and $\mathbf{u}''$, instead of $\mathbf{u}'$ and
$\mathbf{v}''$,
\begin{equation}
\tilde{\mathcal S} \left( { \mathbf{u}''}, \,  \mathbf{u}',T \right)
  = \mathcal S \left( { \mathbf{v}''}, \,  \mathbf{u}',T \right)
   + i \hbar \mathbf{u}'' \mathbf{v}'' \, ,
\label{Tlegen}
\end{equation}
and satisfies the relations
\begin{equation}
\mathbf{v}'' =
   -\frac{i}{\hbar}
   \frac{\partial \tilde{\mathcal S}}{\partial \mathbf{u}''}
   \quad \mathrm{and} \quad
\mathbf{v}' =
   \frac{i}{\hbar}
   \frac{\partial \tilde{\mathcal S}}{\partial \mathbf{u}'}  \, .
\label{stilpartial}
\end{equation}
With these properties in mind we define a dual representation
$\tilde{\mathrm K} \left( { \mathbf{u}''}, \, \mathbf{u}',T \right)$
for the propagator ${\mathrm K} \left( { \mathbf{v}''}, \,
\mathbf{u}',T \right)$ by
\begin{eqnarray}
\tilde{\mathrm K} \left( { \mathbf{u}''}, \,  \mathbf{u}',T \right)
&=&
   \frac{1}{2 \pi }
   \int_C{
      \mathrm{K} \left( { \mathbf{v}''}, \,  \mathbf{u}' ,T\right)
      e^{  - \mathbf{u}'' \mathbf{v}'' }
      \mathrm{d}^2 \mathbf{v}''
   } , \label{ktil} \\ \nonumber \\
\mathrm{K} \left( { \mathbf{v}''}, \,  \mathbf{u}',T \right)  &=&
   \frac{1}{2 \pi }
   \int_{\tilde{C}}{
      \tilde{\mathrm K} \left( { \mathbf{u}''}, \,  \mathbf{u}',T \right)
      e^{   \mathbf{u}'' \mathbf{v}'' }
      \mathrm{d}^2 \mathbf{u}''
   }, \label{ktilinv}
\label{eq8}
\end{eqnarray}
where $C$ and $\tilde{C}$ are convenient paths that, as specified in
\cite{prl}, are chosen in such a way that Eqs. (\ref{ktil}) and
(\ref{ktilinv}) are a Laplace and a Mellin transform, respectively.
The analogy between these two expressions and the corresponding
coordinate and momentum representations is not complete. This is
because, while $\mathrm{K}( { \mathbf{v}''}, \, \mathbf{u}',T)$ is
the quantum propagator, $\tilde{\mathrm{K}}( { \mathbf{u}''}, \,
\mathbf{u}',T)$ does not seem to correspond to an explicit quantum
matrix element. It is a mathematical artifice that allows for the
continuation of the propagator in an auxiliary phase space, rather
than a quantity with a direct physical interpretation.

In order to obtain a semiclassical formula for $\tilde{\mathrm K}
\left( { \mathbf{u}''}, \,  \mathbf{u}',T \right)$, we insert Eq.
(\ref{eq1}) into (\ref{ktil}),

\begin{equation}
\tilde{\mathrm K}_{\mathrm{SC}} \left( { \mathbf{u}''}, \,
\mathbf{u}',T \right)  =
   \frac{1}{2 \pi }
   \int_C{~e^{
      \frac{i}{\hbar} \mathcal{S} \left( { \mathbf{v}''}, \,  \mathbf{u}' ,T\right) +
      \frac{i}{\hbar} \mathcal{G} \left( { \mathbf{v}''}, \,  \mathbf{u}' ,T\right)
      - \frac{i}{2}\sigma_{\mathbf{vv}}  -\frac{1}{2} \ln |\det \mathrm{M_{vv}}|
      -\mathbf{u}'' \mathbf{v}''}
      \mathrm{d}^2 \mathbf{v}''
   } . \label{ktilsc}
\end{equation}
Rigorously, Eq.~(\ref{ktilsc}) says that to calculate
$\tilde{\mathrm K}_{\mathrm{SC}}$ for a set of parameters
$\mathbf{u}'', \,  \mathbf{u}'$ and $T$, we need to calculate the
contribution of the trajectory beginning at
$\mathbf{u}(0)=\mathbf{u}'$ and ending at
$\mathbf{v}(T)=\mathbf{v}''$, and sum over all $\mathbf{v}''$ lying
in the path ${C}$. Notice that, for each trajectory, the value of
the variable $\mathbf u$ at time $T$ is function of $\mathbf{u}'$,
$\mathbf{v}''$ and $T$, namely,
$\mathbf{u}(T)\equiv\mathbf{u}(\mathbf{v}'', \mathbf{u}', T)$. In
the semiclassical limit this integral can be solved by the steepest
descent method~\cite{bleistein}, according to which the critical
value $\mathbf{v}''_c$ of the integration variable satisfies
\begin{equation}
\left.\left\{\frac{\partial}{\partial \mathbf{v}''} \left[
\mathcal{S} + i\hbar
\mathbf{u}''\mathbf{v}''\right]\right\}\right|_{\mathbf{v}''_c} =0
\quad
  \mbox{or} \quad
\mathbf{u}'' =
   \left.\frac{i}{\hbar}\frac{\partial \mathcal{S}}{\mathbf{v}''}\right|_{\mathbf{v}''_c},
\label{crit}
\end{equation}
where we have considered that $\mathcal{G}$ and $\ln |\det
\mathrm{M_{vv}}|$ varies slowly in comparison with $\mathcal{S}$,
since the former is of order $\hbar$ while the later is of order
$\hbar^0$ (see Ref. \cite{Bar01}). Eq.(\ref{crit}) says that the
critical trajectory satisfies $\mathbf{u}(0)=\mathbf{u}'$ and
$\mathbf{u}(T)=\mathbf{u}(\mathbf{v}''_c, \mathbf{u}',
T)=\mathbf{u}''$, i.e., the critical value $\mathbf{v}_c''$ of the
integration variable is equal to $\mathbf{v}(T)$ of a trajectory
satisfying these boundary conditions. This shows that the
integration path $C$ must coincide with (or be deformable into) a
steepest descent path through $\mathbf{v}_c''$. Expanding the
exponent up to second order around this trajectory and performing
the resulting Gaussian integral we obtain
\begin{equation}
\tilde{\mathrm K}_{\mathrm{SC}}^{\mathrm{(2)}} \left( { \mathbf{u}''}, \,  \mathbf{u}',T \right) =
   \sum_{\mathrm{traj.}}
   \sqrt{\frac{1}{|\det \mathbf{\mathrm M _{u v}}|}} \;
   \exp
   \left\{
      \frac{i}{\hbar} \tilde{\mathcal S}
      \left( { \mathbf{u}''}, \,  \mathbf{u}',T \right)
      + \frac{i}{\hbar} \tilde{\mathcal G}
      \left( { \mathbf{u}''}, \,  \mathbf{u}' ,T\right) - \frac{i}{2}
      \sigma_{\mathbf{uv}}
   \right\} \, ,
\label{eq9}
\end{equation}
where, again, the index $^{(2)}$ is used to indicate the method of
integration used. The sum over stationary trajectories was included
because more than one of them may exist. To derive the last
equation, we have also used the result
\begin{equation}
-\det
\left(
   \begin{array}{cc}
     {\mathcal S}_{v_x''v_x''} & {\mathcal S}_{v_x''v_y''} \\
     {\mathcal S}_{v_y''v_x''} & {\mathcal S}_{v_y''v_y''} \\
   \end{array}
\right)=
\hbar^2
\frac{|\det \mathbf{\mathrm M _{u v}}|}
{|\det\mathbf{\mathrm M _{v v}}|}
e^{i({\sigma}_{\mathbf{uv}}-\sigma_{\mathbf{vv}})},
\label{Svv}
\end{equation}
with ${\mathcal S}_{\alpha \beta}\equiv\partial^2{\mathcal
S}/\partial \alpha\partial\beta$, for $\alpha,\beta=v_x''$ or
$v_y''$,  and $\sigma_{\mathbf{uv}}$ is the phase of $\det \mathbf{\mathrm
M _{u v}}$. This last equation can be obtained by considering small
variations of Eq. (\ref{ull}), rearranging the terms so as to write
$\delta \mathbf{u}''$ and $\delta \mathbf{v}''$ as function of
$\delta \mathbf{u}'$ and $\delta \mathbf{v}'$, and comparing with
Eq. (\ref{eq5}).

The new semiclassical propagator $\tilde{\mathrm K}_{\mathrm{SC}}$
is a function of complex classical trajectories satisfying
$\mathbf{u}'=\mathbf{u}(0)$ and $\mathbf{u}'' =\mathbf{u}(T)$.
$\mathbf{\mathrm M _{u v}}$ is given by Eq.~(\ref{eq5}),
$\tilde{\mathcal G}\left( { \mathbf{u}''}, \,\mathbf{u}',T \right) $
is the function $\mathcal{G}$ calculated at the new trajectory, and
$\tilde{\mathcal S} \left( {\mathbf{u}''}, \, \mathbf{u}',T \right)$
is given by Eq.~(\ref{Tlegen}). It is easy to see from
Eq.~(\ref{eq6}) that, when $\det \mathbf{\mathrm M _{v v}}$ is zero,
$\det \mathbf{\mathrm M _{u v}}$ is not. This is a fundamental
property that one has to bear in mind when deriving approximations
for $\mathrm{K}\left( { \mathbf{v}''}, \,  \mathbf{u}' ,T\right)$
inserting $\tilde{\mathrm K}_{\mathrm{SC}}^{\mathrm{(2)}} $ into
Eq.~(\ref{ktilinv}). Three such approximations will be derived in
the next section.

%%%%%%%%%%%%%%%%%%%%%%%%%%%%%%%%%%%%%%%%%%%%%%%%%%%%%%%%%%%%%%%%%%%%
%%%%%%%%%%%%%%%%%%%%%%%%%%%%%%%%%%%%%%%%%%%%%%%%%%%%%%%%%%%%%%%%%%%%
\section{Coherent State Propagator from its Dual Representation}

Replacing Eq. (\ref{eq9}) back into Eq. (\ref{ktilinv}) we obtain
\begin{eqnarray}
\mathrm{K_{SC}} \left( { \mathbf{v}''}, \,  \mathbf{u}',T \right)
&=&
   \frac{1}{2 \pi }
   \int_{\tilde{C}}{~
   e^{ \frac{i}{\hbar}\tilde{\mathcal S}
      \left( { \mathbf{u}''}, \,  \mathbf{u}',T \right)+ \frac{i}{\hbar}
       \tilde{\mathcal G}
      \left( { \mathbf{u}''}, \,  \mathbf{u}' ,T\right) - \frac{i}{2}
      \sigma_{\mathbf{uv}}-  \frac{1}{2} \ln |\det \mathbf{\mathrm M _{u v}}|+
      \mathbf{u}'' \mathbf{v}''  }~ \mathrm{d}^2 \mathbf{u}'' }. \label{ktilinvsc}
\end{eqnarray}
To solve $\mathrm{K_{SC}}$ for the parameters $\mathbf{v}'', \,
\mathbf{u}'$ and $T$, we need to sum the contributions of all
trajectories beginning at $\mathbf{u}'$ and ending at $\mathbf{u}''$
lying in $\tilde{C}$. The saddle point $\mathbf{u}''_c$ of the
exponent satisfies
\begin{equation}
\left.\left\{ \frac{\partial}{\partial \mathbf{u}''} \left[
 \tilde{\mathcal{S}} - i\hbar
\mathbf{u}''\mathbf{v}''\right]\right\}\right|_{\mathbf{u}''_c} =0
\qquad
  \mbox{or} \qquad
\mathbf{v}'' =
   \left.-\frac{i}{\hbar}\frac{\partial \mathcal{S}}{\mathbf{v}''}
   \right|_{\mathbf{u}''_c},
\label{bc}
\end{equation}
which says that the most contributing trajectories are those with
boundary conditions $\mathbf{v} (T)=\mathbf{v}''$ and
$\mathbf{u}(0)=\mathbf{u}'$, exactly as in Eq.~(\ref{eq1}).
Therefore, expanding the exponent up to second order around the
critical trajectory, solving the remaining Gaussian integral, and
using the result (see Eq.~(\ref{b12}) of the appendix)
\begin{equation}
-\det \left(
   \begin{array}{cc}
     \tilde{\mathcal S}_{u_x''u_x''} & \tilde{\mathcal S}_{u_x''u_y''} \\
     \tilde{\mathcal S}_{u_y''u_x''} & \tilde{\mathcal S}_{u_y''u_y''} \\
   \end{array}
\right) \equiv -\det \tilde{\mathrm{S}}_{\mathbf{ u''u''}} =\hbar^2
\frac{|\det\mathbf{\mathrm M_{v v}}| } {|\det\mathbf{\mathrm M_{u
v}}|} e^{i(\sigma_{\mathbf{vv}} - {\sigma_{\mathbf{uv}}})}, \label{eq21}
\end{equation}
we recover Eq. (\ref{eq1}).

Clearly, the connection between the propagators of Eqs. (\ref{eq1})
and (\ref{eq9}) via steepest descent approximation with quadratic
expansion of the exponent works only in the regions where both $\det
\mathbf{\mathrm M _{u v}}$ and $\det \mathbf{\mathrm M _{v v}}$ are
non-zero. Close to caustics, where $\det \mathbf{\mathrm M _{v
v}}=0$, $\tilde{\mathrm K}^{(2)}_{\mathrm{SC}}$ is still well
defined and $\mathrm{ K_{SC}}$ can be obtained by doing the inverse
transform (\ref{ktilinvsc}) but expanding the exponent to at least
third order. There are, however, several ways to handle such an
expansion, depending on how close to the caustic a given stationary
trajectory is. In the next subsections we show how to obtain three
approximate formulas for the propagator:

In Sect.~\ref{regular}, we evaluate Eq. (\ref{ktilinvsc}) by
expanding its integrand up to third order around the stationary
trajectories. As a result we find that each contribution already
present in $\mathrm{ K_{SC}^{(2)}}$ appears multiplied by a
correction term $\mathcal I_R$. This {\it regular formula} for the
semiclassical propagator is good only if the stationary trajectories
are not too close to caustics, so that second and third order terms
contribute to the integral.

In Sec.~\ref{transitional}, we consider the situation where two
contributing solutions are so close each other that, if we used the
regular formula, the contributions would be counted twice. We
therefore perform a {\em transitional} approximation, where the
exponent of (\ref{ktilinvsc}) is expanded  around the trajectory
that lies exactly at the phase space caustic. Since this trajectory
is not generally stationary, this approach works only if the
stationary solutions are sufficiently close to the caustic.

Finally, in Sect.~\ref{uniform}, we derive a {\it uniform
approximation}, which is applicable both near and far from the
caustics but might not be so accurate as the
two previous expressions.

%%%%%%%%%%%%%%%%%%%%%%%%%%%%%%%%%%%%%%%%%%%%%%%%%%%%%%%%%%%%%%%%%%%%
\subsection{Regular Formula}
\label{regular}

The philosophy of the regular approximation is to correct the
contribution of each stationary trajectory by including third order
terms in the expansion of the exponent of Eq.~(\ref{ktilinvsc}).
When this expansion is performed we obtain
\begin{equation}
\mathrm{K_{SC}^{(3)}} \left( { \mathbf{v}''}, \,  \mathbf{u}',\,T \right)  =
%\mathrm{K_{SC}^{(2)}} \left( { \mathbf{v}''}, \,  \mathbf{u}',\,T \right)
\left\{ \sqrt{\frac{1}{|\det \mathbf{\mathrm M _{v v}}|}}~
e^{\frac{i}{\hbar}\mathcal F}\right\} \times \mathcal I_R\left( {
\mathbf{v}''}, \,  \mathbf{u}',\,T \right), \label{eqk}
\end{equation}
where the quantities between brackets are the same as in
Eq.~(\ref{eq1}), and the correction term $\mathcal I_R$ is given by
\begin{equation}
\mathcal{I}_R  =
\sqrt{-\frac{\det\tilde{\mathrm S}_{\mathbf{u}''\mathbf{u}''}}{4\pi^2\hbar^2}}
   \int \mathrm{d}^2 [\delta \mathbf{u}'' ]~
   e^{ \frac{i}{\hbar}
    \left\{ A\delta u_x''^2  + B\delta u_x'' u_y''
    + C\delta u_y''^2 +
     D\delta  u_x''^3 + E\delta  u_x''^2  \delta u_y''
    + F \delta u_y''^2 \delta u_x'' + G\delta u_y''^3
   \right\} }. \label{ia1}
\end{equation}
and
\begin{equation}
\begin{array}{lllllll}
A= \frac{1}{2} \tilde{\mathcal S}_{u_x''u_x''} ,&&
B=\tilde{\mathcal S}_{u_x''u_y''} ,&&
C=\frac{1}{2} \tilde{\mathcal S}_{u_y''u_y''} ,&&\\ \\
D= \frac{1}{6} \tilde{\mathcal S}_{u_x'' u_x'' u_x''} ,&& E=
\frac{1}{2} \tilde{\mathcal S}_{u_x'' u_x'' u_y''} ,&& F=
\frac{1}{2} \tilde{\mathcal S}_{u_x'' u_y'' u_y''}
&&\mathrm{and}\;\; G=\frac{1}{6} \tilde{\mathcal S}_{u_y'' u_y''
u_y''}.
\end{array}
\label{ctes}
\end{equation}
All functions and constants in Eq.~(\ref{eqk}) are calculated at the
critical trajectory. In Eq.~(\ref{ctes}), we define $\tilde{\mathcal
S}_{\alpha \beta \gamma}\equiv(\partial^3\tilde{\mathcal S}/\partial
\alpha\partial\beta\partial \gamma)$ and $\tilde{\mathcal S}_{\alpha
\beta}\equiv(\partial^2\tilde{\mathcal S}/\partial
\alpha\partial\beta)$, for $\alpha,\beta,\gamma=u_x''$ or $u_y''$.
The integration contour of Eq.~(\ref{ia1}) is chosen to coincide
with the steepest descent of the saddle point.

The integral~(\ref{ia1}) has no direct solution. However, it can be
largely simplified in the coordinate system $(\delta u_+,\delta
u_-)$ that diagonalizes the matrix the quadratic terms,
\begin{equation}
\left( \begin{array}{cc} A & B/2 \\ B/2 & C \end{array} \right)
=
 \frac{1}{2} \tilde{\mathrm S}_{\mathbf{u}''\mathbf{u}''}
\;.
\label{quadr}
\end{equation}
Therefore, we perform the change of variables
\begin{equation}
\left[ \begin{array}{c}\delta u_x'' \\ \delta u_y'' \end{array} \right] =
    \frac{1}{B/2(\lambda_- - \lambda_+)}
\left[ \begin{array}{cc} N_+(A - \lambda_-) \quad & -N_-(A - \lambda_+) \\
     N_+ B/2 \quad & -N_- B/2 \end{array} \right]
\left[ \begin{array}{c}\delta u_+ \\ \delta u_- \end{array} \right] \; ,
\label{muz}
\end{equation}
where $N_{\pm}$ are normalization constants and $\lambda_{\pm}$ are
eigenvalues of $ \frac{1}{2} \tilde{\mathrm
S}_{\mathbf{u}''\mathbf{u}''}$,
\begin{equation}
N_{\pm} =  \sqrt{ \left( B/2 \right)^2 + \left( A - \lambda_{\pm}
\right)^2 } \qquad \mathrm{and} \qquad
 \lambda_{\pm} = \frac{\mathrm{tr} \, \mathrm{\tilde{\mathrm S}_{\mathbf{u}''\mathbf{u}''}}}{4}
\left\{ 1 \pm \sqrt{1 - 4 \frac{\det  \tilde{\mathrm S}_{\mathbf{u}''\mathbf{u}''}}
{({\mathrm{tr} \, \tilde{\mathrm S}_{\mathbf{u}''\mathbf{u}''})}^2 }} \right\} \, .
\label{norm}
\end{equation}
In the new variables Eq.~(\ref{ia1}) becomes
\begin{equation}
\mathcal{I}_R =
\sqrt{-\frac{{\lambda_+ \lambda_-}}{\pi^2\hbar^2}}
\int \mathrm{d}[\delta u_+] \mathrm{d}[\delta u_-] ~
e^{
 \frac {i}{\hbar}\left\{  \lambda_+\delta u_+^2 + \lambda_-\delta u_-^2
+ D'\delta u_+^3 + E'\delta u_+^2 \delta u_- + F'\delta u_+ \delta u_-^2 + G'\delta u_-^3
 \right\}}
\, ,
\label{G3Te}
\end{equation}
where the new coefficients, $D', \, E' ,\,F'$ and $G'$, are
combinations of those in Eq.~(\ref{ctes}). Our final formula depends
just on $G'$, which amounts to
\begin{equation}
G' = \left( \frac{N_-}{\lambda_+ - \lambda_-}
\right)^3
\left[
\left( \frac{A-\lambda_+}{B/2} \right)^3 D +
\left( \frac{A-\lambda_+}{B/2} \right)^2 E  +
\left( \frac{A-\lambda_+}{B/2} \right) F + G  \right] .
\label{gl}
\end{equation}

According to Eqs.~(\ref{eq21}) and (\ref{quadr}), when $\det
\mathrm{M_{\mathbf{vv}}}\rightarrow 0$, $\det\tilde{\mathrm
S}_{\mathbf{u}''\mathbf{u}''}$ also tends to zero, causing the
breaking down of the quadratic approximation. However, in terms of
the variables $\delta u_+$ and $\delta u_-$, we see that
$\det\tilde{\mathrm S}_{\mathbf{u}''\mathbf{u}''}$
($=4\lambda_+\lambda_-$) goes to zero in a particular way: while
$\lambda_-$ vanishes, $\lambda_+$ generally remains finite.
Therefore Eq.~(\ref{G3Te}) is always of a gaussian type integral in
the variable $\delta u_+$, since we are still able to neglect third
order terms in this direction. Solving the integral in $\delta u_+$
leads to
\begin{equation}
\mathcal{I}_R  \approx   \sqrt{\frac{-i \lambda_-}{\pi\hbar}}
\int{ \mathrm{d}[\delta u_-] ~e^{ \frac {i}{\hbar}
\left\{ \lambda_- \delta u_-^2 + G'  \delta u_-^3 \right\} }} \, .
\label{G3Tereau}
\end{equation}
Now we perform a last changing of variables, $t = \left(
\frac{3G'}{\hbar}\right)^{1/3} \left[\delta u_- +
\frac{\lambda_-}{3G'} \right]$, and obtain
\begin{equation}
\mathcal{I}_R \approx
~2\sqrt\pi ~\bar w~e^{ \frac{2}{3}\bar w^6}~
\mathrm{f_{i}} (\bar w^4) \, ,
\label{G3Tereaupfim}
\end{equation}
where $\bar
w=\frac{\left(-i\lambda_-/\hbar\right)^{1/2}}{\left(3G'/\hbar\right)^{1/3}}$
and $\mathrm{f_i}(w)$ is given by
\begin{eqnarray}
\mathrm{f_{i}}(w)
=
\frac{1}{2\pi}\int_{C_i} \mathrm{d} t
\exp{ \left\{
i \left[ wt  + \frac{1}{3} t^3 \right] \right\}} ,
\label{int7}
\end{eqnarray}
for $\mathrm i=1,2,3$. The index $\mathrm i$ refers to three
possible paths of integration $C_i$, giving rise to three different
Airy's functions (see Ref. \cite{bleistein}). Rigorously the choice
of the path should be done according to Cauchy's Theorem, after all
the path to be used has to be obtained by a deformation of the
original contour of integration. In practice, however, it might be
very difficult to find the correct path in this way, and we have to use
physical criteria to justify the choice of $C_i$.

Inserting (\ref{G3Tereaupfim}) into Eq.~(\ref{eqk}) and considering
the existence of more than one critical trajectory, we finally find
the regular formula
\begin{equation}
\mathrm{K_{SC}^{(3)}} \left( { \mathbf{v}''}, \,  \mathbf{u}',\,T \right)  =
\sum_{\mathrm{traj.}}
\left\{ \left[
\sqrt{\frac{1}{|\det \mathrm{M}_{\mathbf{vv}}|}} ~
e^{\frac{i}{\hbar}\mathcal F\left( { \mathbf{v}''}, \,  \mathbf{u}',\,T \right)}
\right]
\times
\left[2\sqrt\pi ~\bar w~e^{ \frac{2}{3}\bar w^6}~
\mathrm{f_{i}} (\bar w^4) \right]\right\}.
\label{regform}
\end{equation}

In this equation, each stationary trajectory gives a contribution
which is that of the quadratic approximation multiplied by a
correction factor $\mathcal I_R$ that depends only on the parameter
$\bar w$. Close to a caustic $\lambda_-$ is very small but $G'$
(generally) remains finite. Exactly at the caustic $|\bar w|$ is
zero, getting larger and larger as we move away from it. Therefore
we expect that $\mathcal I_R$ should go to 1 as $|\bar w|$ goes to
infinity, since the regular expression should recover
$\mathrm{K_{SC}^{(2)}}$ in this limit. To verify this assertion, we
look at the asymptotic formulas for the Airy's functions
\cite{abra},
\begin{equation}
%\left.
\begin{array}{l}
\mathrm{f_1}(w) \sim \frac{1}{2\sqrt{\pi}} w^{-1/4} e^{-\frac{2}{3}w^{3/2}},\\
\mathrm{f_2}(w) \sim \frac{-i}{2\sqrt{\pi}} w^{-1/4} e^{\frac{2}{3}w^{3/2}},\\
\mathrm{f_3}(w) \sim \frac{i}{2\sqrt{\pi}} w^{-1/4} e^{\frac{2}{3}w^{3/2}}.
\end{array}
%\right\}|w|\rightarrow\infty, \;|\mathrm{phase\;of\;}w |.
\label{airyass}
\end{equation}
Using these expressions in Eq.~(\ref{G3Tereaupfim}), we see that
only $\mathrm{f_1}(w)$ produces the desired asymptotic result,
indicating that this is the proper choice of Airy function far from
the caustic. However, this is so only because we have taken the
principal root in the definition of $\bar w$. As the physical
results should not depend on the arbitrariness of branches in the
complex plane, the choice of a different root would lead to a
different path $C_i$, so that physical results remain the same. A
careful discussion about this point can be found in \cite{parisio2}.

Exactly at the caustic, $\bar w= 0$, the regular formula becomes
\begin{equation}
\mathrm{K_{SC}^{(3)}} \left( { \mathbf{v}''}, \,  \mathbf{u}',\,T \right)  =
\sqrt{\frac{i\hbar\pi}{\lambda_+\left(\det \mathrm{M}_{\mathbf{uv}}\right)}}
\left(\frac{\hbar}{3G'}\right)^{1/3}\mathrm{f_{i}} (0)~
e^{\frac{i}{\hbar}\left[\mathcal S
+\mathcal G\right]}
 \qquad (\mathrm{with}\;\bar w =0),
\label{regformpsc}
\end{equation}
where the sum was excluded because the critical trajectories
coalesce at this point.

%%%%%%%%%%%%%%%%%%%%%%%%%%%%%%%%%%%%%%%%%%%%%%%%%%%%%%%%%%%%%%%%%%%%
\subsection{Transitional Formula}
\label{transitional}

Each contribution to the semiclassical propagator calculated in the
last section (as well as those of Eq.~(\ref{eq1})) has information
about the critical trajectory plus its vicinity. If two trajectories
are very close each other, like in the vicinity of a phase space
caustic, their regions of influence might overlap. The regular
formula cannot be used in these situations, since it assumes that
the trajectories can still be counted independently. To find an
approximation for $\mathrm{K} \left( { \mathbf{v}''}, \,
\mathbf{u}',\,T \right)$ valid in this scenario, we shall perform
the integral~(\ref{ktilinvsc}) expanding the exponent about the
(non-stationary) trajectory corresponding to the phase space caustic
itself, defined by the point $\bar{\mathbf{u}}''$ where $|\det
\tilde{\mathrm S}_{\mathbf{u}''\mathbf{u}''}|$ and therefore $|\det
\mathbf{\mathrm M _{v v}}|$ [see Eq. (\ref{eq21})] are zero.
Evaluating this single contribution to third order should be
equivalent to include and sum over each stationary trajectory.

The condition $\det \tilde{\mathrm S}_{\mathbf{u}''\mathbf{u}''} =
0$ leads to the PSC trajectory that begins at
$\mathbf{u}(0)\equiv{\mathbf{u}}'$ and
$\mathbf{v}(0)\equiv{\mathbf{v}}'$, and ends at
$\mathbf{u}(T)\equiv\bar{\mathbf{u}}''$ and
$\mathbf{v}(T)\equiv\bar{\mathbf{v}}''$, where $\bar{\mathbf{v}}''$
is assumed to be close to ${\mathbf{v}}''$. Expanding the exponent
of Eq. (\ref{ktilinvsc}) up to third order around this new
trajectory yields
\begin{equation}
\mathrm{K_{SC}^{PSC}} \left( { \mathbf{v}''}, \,  \mathbf{u}',\,T \right)  =
   \left(\det \mathbf{\mathrm M _{u v}}\right)^{-1/2}
   e^{
      \frac{i}{\hbar}
         \left\{
         {\mathcal S}
      \left( { \bar{\mathbf{v}}''}, \,  \mathbf{u}',T \right)
      +  {\mathcal G}
      \left( { \bar{\mathbf{v}}''}, \,  \mathbf{u}',T \right)
%      - \frac{\hbar}{2} \tilde{\sigma}
      -i\hbar \bar{\mathbf{u}}''(\mathbf{v}'' - \bar{\mathbf{v}}'')
   \right\}} ~  \mathcal{I}_T,
\label{eqkt}
\end{equation}
where
\begin{equation}
\mathcal{I}_T  =
   \frac{1}{2\pi}
   \int \mathrm{d}^2 [\delta \mathbf{u}'' ]~
   e^{ \frac{i}{\hbar}
    \left\{ X  \delta u_x'' + Y \delta u_y''
    + A\delta u_x''^2  + B\delta u_x'' u_y''
    + C\delta u_y''^2 +
     D\delta  u_x''^3 + E\delta  u_x''^2  \delta u_y''
    + F \delta u_y''^2 \delta u_x'' + G\delta u_y''^3
   \right\} }, \label{ia1t}
\end{equation}
with $X=\partial\tilde{\mathcal S}/\partial{u_x''} - i \hbar v_x''$
and $Y= \partial\tilde{\mathcal S}/\partial{u_y''} - i \hbar v_y''$.
The functions appearing in Eq.~(\ref{eqkt}) and all the coefficients
are calculated at the PSC trajectory.

We solve Eq.~(\ref{ia1t}) using the same technique described in the
last section, with the use of the transformation~(\ref{muz}).
However, as we deal with the PSC trajectory, $\lambda_+ =  A+C $ and
$\lambda_- = 0$. The integral $\mathcal{I}_T$ becomes
\begin{equation}
\mathcal{I}_T = \frac{1}{2\pi}\int  \mathrm{d}[\delta u_+] \mathrm{d}[\delta u_-]
e^{
\frac {i}{\hbar} \left\{ a\delta u_+ + b\delta u_- + \lambda_+ \delta u_+^2 + D'\delta
u_+^3 + E'\delta u_+^2\delta  u_- + F'\delta u_+\delta u_-^2 + G'\delta u_-^3  \right\}},
\label{int}
\end{equation}
where the only coefficients that appear in the final formula are
\begin{equation}
a=-\left(\frac{N_+}{\lambda_+-\lambda_-}\right)\left[
\left(\frac{A-\lambda_-}{B/2}\right)X+Y\right],
\quad
b=\left(\frac{N_-}{\lambda_+-\lambda_-}\right)\left[
\left(\frac{A-\lambda_+}{B/2}\right)X+Y\right]
\end{equation}
and $G'$, given by Eq.~(\ref{gl}).

The integral over $\delta u_+$ can be performed neglecting terms of
third order. We obtain
\begin{equation}
\mathcal{I}_T =
 \frac{1}{2\pi}
\sqrt{\frac{i\pi\hbar}{\lambda_+}}
~e^{
-\frac{i}{\hbar} \frac{a^2}{4 \lambda_+} }
\int \mathrm{d} [\delta u_-]
\exp{ \left\{
\frac {i}{\hbar} \left[ b\delta u_-  + G' \delta u_-^3 \right] \right\}}.
\label{int5}
\end{equation}
By setting $t=\left( \frac{3G'}{\hbar}\right)^{1/3} \delta u_-$, the
last equation can be written as
\begin{equation}
\mathcal{I}_T =
\sqrt{\frac{i\pi\hbar}{\lambda_+}}
~e^{
-\frac{i}{\hbar} \frac{a^2}{4 \lambda_+} }
\left( \frac{\hbar}{3G'}\right)^{1/3}
\mathrm{f_{i}}(\tilde w),
\label{int6}
\end{equation}
where $\tilde w =\frac{b/\hbar}{(3G'/\hbar)^{1/3}}$ and the function
$\mathrm{f_{i}}(w)$ refers to the Airy's functions~(\ref{int7}).
Finally, we write the transitional formula by combining Eq.
(\ref{int6}) with Eq. (\ref{eqkt}),
\begin{equation}
\mathrm{K_{SC}^{PSC}} \left( { \mathbf{v}''}, \,  \mathbf{u}' ,T\right) =
   \sqrt{\frac{i\hbar\pi}{\lambda_+\left(\det \mathbf{\mathrm M _{u v}}\right)}}
~\left( \frac{\hbar}{3G'}\right)^{1/3}
e^{
-\frac{i}{\hbar} \frac{a^2}{4 \lambda_+} }
~\mathrm{f_{i}}\left(\tilde w\right)~e^{
\frac{i}{\hbar} [\mathcal{ S+G}]-i\hbar\bar{\mathbf{u}}''(\mathbf{v}'' -
\bar{\mathbf{v}}'')}.
\label{eqkf}
\end{equation}
Equation~(\ref{eqkf}) depends on the PSC trajectory, which satisfies
$\mathbf{u}(0)=\mathbf{u}'$ and $\mathbf{v}(T)=\bar{\mathbf{v}}''$,
and is valid only if $\bar{\mathbf{v}}''$ is close to
${\mathbf{v}''}$.

Far from the caustic Eq.~(\ref{eqkf}) does not make sense, since the
PSC trajectory becomes completely different from the actual
stationary trajectories. On the other hand, when the propagator is
calculated exactly at the PSC, Eqs.~(\ref{eqkf}) and (\ref{regform})
should furnish the same result. This can be verified by setting
$\bar{\mathbf{v}}''={\mathbf{v}}''$ and $a=b=\tilde w=0$ in
Eq.~(\ref{eqkf}), which reduces directly to Eq.~(\ref{regformpsc}).

%%%%%%%%%%%%%%%%%%%%%%%%%%%%%%%%%%%%%%%%%%%%%%%%%%%%%%%%%%%%%%%%%%%%%%
\subsection{Uniform Formula}
\label{uniform}

The regular formula is good as long as one is not too close to a
phase space caustic, whereas the transitional formula is good only
very close to it. In either cases the expressions we derived cannot
be used everywhere in the space spanned by the parameters
$\mathbf{u}'$, $\mathbf{v}''$ and $T$. The uniform approximation
provides such a global formula \cite{uniform}. The basic idea is to
map the argument of the exponential in (\ref{ktilinvsc}) into a
function having the same structure of saddle points as the original
one, i.e., two saddle points that may coalesce on the phase space
caustic depending on a given parameter.

In order to simplify our calculation, we shall use the variables
$u''_+$ and $u''_-$,  instead of the original $u''_x$ and $u''_y$
[see Eq. (\ref{muz})]. In these variables the exponent of
Eq.~(\ref{ktilinvsc})
\begin{equation}
E\left( { \mathbf{u}''}, \,  \mathbf{u}',T \right)  =
   { \frac{i}{\hbar}\tilde{\mathcal S}
      \left( { \mathbf{u}''}, \,  \mathbf{u}',T \right)+
      \frac{i}{\hbar}\tilde{\mathcal G}
      \left( { \mathbf{u}''}, \,  \mathbf{u}' ,T\right) - \frac{i}{2}
      \sigma_{\mathbf{uv}}-  \frac{1}{2} \ln |\det \mathbf{\mathrm M _{u v}}|
      + \mathbf{u}'' \mathbf{v}''  },
\label{exp}
\end{equation}
becomes
\begin{equation}
\mathcal E( u''_+,u''_-) \equiv
E\left[ { \mathbf{u}''(u''_+,u''_-)}, \,  \mathbf{u}',T \right],
\end{equation}
where we omit the dependence on the variables $\mathbf{u}'$ and $T$
because they are not being integrated. The integral
(\ref{ktilinvsc}) then becomes
\begin{eqnarray}
   \frac{1}{2 \pi }
   \int{~
   e^{ \mathcal E\left( u''_+,u''_-\right)}~ \mathrm{d}u_+''\mathrm{d}u_-'' }.
   \label{iu1}
\end{eqnarray}

Since the main contributions to this integral comes from the
neighborhood of the saddle points, we can map the exponent $\mathcal
E( u''_+,u''_-)$ into a new function $N(x,y)$, where $x=x(u''_+)$
and $y=y(u''_-)$. We restrict ourselves to the case where there are
only two critical points, $\mathbf u''_1=( u''_+,u''_-)_1$ and
$\mathbf u''_2=( u''_+,u''_-)_2$, which, depending on the parameters
$\mathbf u'$ and $T$, may coalesce at the phase space caustic. Then
\begin{eqnarray}
   \frac{1}{2 \pi }
   \int{~
   e^{ \mathcal E\left( u''_+,u''_-\right)}~ \mathrm{d}u_+''\mathrm{d}u_-'' }
 \approx \frac{1}{2 \pi }
   \int{~ J(x,y,) e^{N(x,y)}~ \mathrm{d}x \mathrm{d}y}.
   \label{iu2}
\end{eqnarray}
The simplest function with these properties is
\begin{equation}
N(x,y)=\mathcal A-\mathcal By+\frac{y^3}{3}+\mathcal Cx^2, \label{n}
\end{equation}
where $\mathcal A$, $\mathcal B$ and $\mathcal C$ may depend on
$\mathbf u'$ and $T$. The mapping requires that the saddle points of
$N(x,y)$, which are $(0,\pm\sqrt{\mathcal B})$, coincide with the
critical points $\mathbf u''_{1,2}$:
\begin{equation}
\begin{array}{l}
\mathcal E( { \mathbf{u}''_1})\equiv \mathcal E_1 =
N(0,\sqrt{\mathcal B}) =
\mathcal A- \frac{2}{3} \mathcal B^{3/2} , \\
\mathcal E( { \mathbf{u}''_2})\equiv \mathcal E_2 =
N(0,-\sqrt{\mathcal B}) = \mathcal A+ \frac{2}{3} \mathcal B^{3/2} ,
\end{array}
\end{equation}
implying that
\begin{equation}
\mathcal A=\frac{1}{2}(\mathcal E_1+\mathcal E_2)
\quad \mathrm{and} \quad
\mathcal B=\left[\frac{3}{4}(\mathcal E_2-\mathcal E_1)\right]^{2/3}. \label{ab}
\end{equation}
Another condition required to validate the method is to impose the
equivalence between the vicinity of critical points of $N(x,y)$ and
$\mathcal E (u''_+,u''_-)$,
\begin{equation}
\left.\left\{\delta N+\frac{1}{2} \delta^2N + \frac{1}{6} \delta^3N
+\ldots\right\} \right|_{(0,\pm\sqrt{\mathcal B})}=
\left.\left\{\delta \mathcal E+\frac{1}{2} \delta^2\mathcal E +
\frac{1}{6}\delta^3\mathcal E + \ldots\right\}\right|_{\mathbf
u''_{1,2}}. \label{vic}
\end{equation}
This equation allows us to find how to transform an arbitrary
infinitesimal vector $(\delta u''_+,\delta u''_-)$ into $(\delta
x,\delta y)$, around the critical points. It provides, therefore,
information about the Jacobian $J(x,y)$ of the transformation
calculated at the critical points, namely, $J_1\equiv
J(0,\sqrt{\mathcal B})$ and $J_2\equiv J(0,-\sqrt{\mathcal B})$.

As the first derivatives of $\mathcal E$ and $N$ vanish at the
critical points, Eq.~(\ref{vic}) implies that
\begin{equation}
\frac{1}{2}
\left( \begin{array}{ll} \delta x & \delta y \end{array}\right)
\left.\left(\begin{array}{cc}
\frac{\partial^2 N}{\partial x^2} + \frac{1}{3}
\frac{\partial^3 N}{\partial x^3} \delta x
&
\frac{\partial^2 N}{\partial x\partial y}+
\frac{\partial^3 N}{\partial y\partial x^2} \delta x
\\
\frac{\partial^2 N}{\partial y\partial x} +
\frac{\partial^3 N}{\partial x\partial y^2} \delta y
&
\frac{\partial^2 N}{\partial y^2} +
\frac{1}{3} \frac{\partial^3 N}{\partial y^3} \delta y
\end{array}\right)\right|_{(0,\pm\sqrt{\mathcal B})}
\left( \begin{array}{ll} \delta x \\ \delta y \end{array}\right)
\end{equation}
should be equal to
\begin{equation}
\frac{1}{2} \left( \begin{array}{ll} \delta u''_+ & \delta u''_-
\end{array}\right) \left.\left(\begin{array}{cc} \frac{\partial^2
\mathcal E }{\partial {u''_+}^2} + \frac{1}{3}\frac{\partial^3
\mathcal E}{\partial {u''_+}^3}\delta u''_+ & \frac{\partial^2
\mathcal E}{\partial u''_+\partial u''_-}+
\frac{\partial^3 \mathcal E }{\partial {u''_-}\partial {u''_+}^2}
\delta u''_+\\
\frac{\partial^2 \mathcal E }{\partial u''_-\partial u''_+} +
\frac{\partial^3 \mathcal E }{\partial {u''_+}\partial {u''_-}^2}
\delta u''_-& \frac{\partial^2 \mathcal E }{\partial {u''_-}^2}+
\frac{1}{3}\frac{\partial^3 \mathcal E }{\partial {u''_-}^3}\delta
u''_-
\end{array}\right)\right|_{\mathbf u''_{1,2}}
\left( \begin{array}{ll} \delta u''_+ \\ \delta u''_- \end{array}\right).
\end{equation}
Writing $\delta u''_+=a_+\delta x$ and $\delta u''_-=a_-\delta y$
this equality results in
\begin{equation}
\begin{array}{l}
\left\{\left[\frac{\partial^2 \mathcal E }{\partial {u''_+}^2} +
\frac{1}{3}\frac{\partial^3 \mathcal E}{\partial {u''_+}^3}
\left(a_+\delta x\right)\right]a_+^2\right\}_{\mathbf u''_{1,2}}
=2\mathcal C,
\\
\left\{\left[ \frac{\partial^2 \mathcal E}{\partial u''_+\partial
u''_-}+ \frac{\partial^3 \mathcal E }{\partial {u''_-}\partial
{u''_+}^2} \left(a_+\delta x\right)\right]a_+a_-\right\}_{\mathbf
u''_{1,2}}=0,
\\
\left\{\left[\frac{\partial^2 \mathcal E }{\partial u''_-\partial
u''_+} + \frac{\partial^3 \mathcal E }{\partial {u''_+}\partial
{u''_-}^2} \left(a_-\delta y\right)\right]a_+a_-\right\}_{\mathbf
u''_{1,2}}=0,
\\
\left\{ \left[ \frac{\partial^2 \mathcal E }{\partial {u''_-}^2}+
\frac{1}{3}\frac{\partial^3 \mathcal E }{\partial
{u''_-}^3}\left(a_-\delta y\right)\right]a_-^2\right\}_{\mathbf
u''_{1,2}} = \pm2\sqrt{\mathcal B}+ \frac{2}{3} \delta y.
\end{array} \label{comp2}
\end{equation}
In the limit of small $\hbar$, $\mathcal G$ and $\det
\mathrm{M_{\mathbf{uv}}}$ vary slowly in comparison with $\mathcal
S$ and the first and last of equations (\ref{comp2}) become,
respectively,
\begin{equation}
\begin{array}{l}
\frac{i}{\hbar}\left.\left\{\left[\lambda_+ + D'\left(a_+\delta x\right)
\right]a_+^2\right\}\right|_{\mathbf u''_{1,2}}=\mathcal C , \\
\frac{i}{\hbar}\left.\left\{\left[ \lambda_-+ G'\left(a_-\delta y\right)
\right]a_-^2\right\}\right|_{\mathbf u''_{1,2}}= \pm\sqrt{\mathcal B}+
\frac{1}{3} \delta y .
\end{array}
\label{comp}
\end{equation}
Moreover, the second and third (\ref{comp2}) imply that $E'=F'=0$.
We emphasize that $D'$, $E'$, $F'$ and $G'$ are the same
coefficients as those of Sec.~\ref{regular}.

Eqs.~(\ref{comp}) can be solved if we neglect the terms containing
$\delta x$ and $\delta y$. We find
\begin{equation}
\left.\left(a_+\right)\right|_{\mathbf u''_{1,2}}=
\sqrt\frac{-i\hbar \mathcal
C}{\left.\left(\lambda_+\right)\right|_{\mathbf u''_{1,2}}} \qquad
\mathrm{and} \qquad \left.\left(a_-\right)\right|_{\mathbf
u''_{1,2}}=\sqrt\frac{\mp i\hbar \sqrt{\mathcal
B}}{\left.\left(\lambda_-\right)\right|_{\mathbf u''_{1,2}}},
\end{equation}
so that the Jacobian at the saddle points becomes
\begin{equation}
J_{1,2}=\left.\left(a_+a_-\right)\right|_{\mathbf u''_{1,2}} =
\sqrt{\frac{\mp \hbar^2 \mathcal C \sqrt{\mathcal B}}
{\left.\left(\lambda_+\lambda_-\right)\right|_{\mathbf u''_{1,2}}}}.
\end{equation}
The full Jacobian can therefore be conveniently written in the
vicinity of the saddle points as
\begin{equation}
J(x,y)=J(y)=\frac{1}{2} \left( J_1 + J_2\right) -
\frac{y}{2\sqrt{\mathcal B}}\left( J_2 - J_1\right),
\end{equation}
and the uniform approximation for the propagator becomes
\begin{equation}
\mathrm{K_{SC}^{UN}} (\mathbf{v}'', \mathbf{u}',T)=
\frac{1}{2\pi}\int J(x,y) ~e^{\mathcal A- \mathcal
By+y^3/3+\mathcal Cx^2}dxdy. \label{equn}
\end{equation}
Performing the integral over $x$ we obtain the final expression
\begin{eqnarray}
\mathrm{K_{SC}^{UN}} (\mathbf{v}'', \mathbf{u}',T)=
i\sqrt\pi~e^{\mathcal A}
\left\{
\left(\frac{g_2-g_1}{\sqrt \mathcal B}\right)\mathrm{f'_i}(\mathcal B)+
(g_1+g_2)\mathrm{f_i}(\mathcal B)
\right\}, \label{uform}
\end{eqnarray}
where $\mathrm{f_i}$ is given by Eq.~(\ref{int7}) and
\begin{equation}
g_{1,2}=\sqrt{\frac{\pm\hbar^2  \sqrt{\mathcal B}}
{\left.\left(4\lambda_+\lambda_-\right)\right|_{\mathbf u''_{1,2}}}}=
\sqrt{\mp  \sqrt{\mathcal B}\left.\left( \frac{\det\mathrm{M_\mathbf{uv}}}
{\det \mathrm{M_\mathbf{vv}}} \right)\right|_{\mathbf u''_{1,2}}}. \label{g1g2}
\end{equation}

Eq.~(\ref{uform}) is the uniform formula for the two-dimensional
coherent state propagator. As in Sects.~\ref{regular} and
\ref{transitional}, the determination of the proper path of
integration $C_i$ is done by physical criteria.

Eq.~(\ref{g1g2}) shows us how the singularity in the coalescence
point is controlled. When $\det \mathrm{M_\mathbf{vv}}$ goes to
zero, the difference between $\mathcal E_1$ and $\mathcal E_2$ also
vanishes, so that the quotient $\sqrt\mathcal B/\det
\mathrm{M_\mathbf{vv}}$ [see also Eq.~(\ref{ab})] remains finite.
Notice, however, that this fraction might become extremely fragile
close to a caustic, because both numerator and denominator go to
zero. Exactly at the caustic we can return to the second of
Eqs.~(\ref{comp}) to find the correct value of $a_-$:
\begin{equation}
a_-^{PSC} = \left( \frac{-i\hbar}{3{G'}}\right)^{1/3}
\Longrightarrow
J_{PSC} = {\left(\frac{-i\hbar \mathcal C}{\lambda_+}\right)^{1/2}}
{\left(\frac{-i\hbar}{3{G'}}\right)^{1/3}}.
\label{o3}
\end{equation}
One should also remember that, if $\hbar$ is not sufficiently small,
the derivatives of $\mathcal G$ and $\det \mathrm{M_\mathbf{vv}}$
may become important, specially when $\lambda_-\rightarrow 0$.

It is interesting to check that the uniform approximation
(\ref{uform}) recovers the quadratic approximation away from the
caustics, i.e., in the limit $\mathcal B \rightarrow \infty$.
According to Eqs.~(\ref{airyass}) we find that, for large $w$,
\begin{equation}
%\left.
\begin{array}{l}
w^{-1/2}\mathrm{f'_1}(w) \sim \frac{-1}{2\sqrt{\pi}}
w^{-1/4} e^{-\frac{2}{3}w^{3/2}},\\
w^{-1/2}\mathrm{f'_2}(w)\sim \frac{-i}{2\sqrt{\pi}}
w^{-1/4} e^{\frac{2}{3}w^{3/2}},\\
w^{-1/2}\mathrm{f'_3}(w) \sim \frac{i}{2\sqrt{\pi}}
w^{-1/4} e^{\frac{2}{3}w^{3/2}}.
\end{array}
\label{airyassl}
\end{equation}
Inserting Eqs.~(\ref{airyass}) and (\ref{airyassl}) into the uniform
approximation results in
\begin{equation}
\mathrm{K_{SC}^{UN}} (\mathbf{v}'', \mathbf{u}',T) \approx
\left\{\begin{array}{ll} -ig_2e^{\mathcal A-\frac{2}{3}\mathcal
B^{3/2}}\mathcal B^{-1/4},&
\mathrm{by\;using\;f_1}\\
-g_1e^{\mathcal A+\frac{2}{3}\mathcal B^{3/2}}\mathcal B^{-1/4},&
\mathrm{by\;using\;f_2}\\
g_1e^{\mathcal A+\frac{2}{3}\mathcal B^{3/2}}\mathcal B^{-1/4},&
\mathrm{by\;using\;f_3}
\end{array}\right. .
\end{equation}
It's easy to see that using the contour $C_1+C_2$ we find
$|\mathrm{K_{SC}^{UN}}|=|\mathrm{K_{SC}^{(2)}}|$.

Another way to arrive at the same conclusion is as follows: if
$\mathbf{u}''_1$ and $\mathbf{u}''_2$ are not close each other, we
can individually evaluate the contribution of each one through the
second order saddle point method and sum the contributions at the
end. Starting from Eq.~(\ref{equn}) we get
\begin{equation}
\begin{array}{lll}
\mathrm{K_{SC}^{UN}} (\mathbf{v}'', \mathbf{u}',T)
&=&
\frac{-i}{2\sqrt{\pi}}\int J(y) e^{\mathcal A- \mathcal By+y^3/3}dy
\\
&=& \frac{-i}{2\sqrt{\pi}}
\sum_{y_0=\pm\sqrt\mathcal B}
\left\{ J(y_0) e^{\mathcal A-\mathcal By_0+y^3_0/3} \int e^{y_0(y-y_0)^2}dy\right\}
\\
&=&
\frac{i\hbar~e^{\mathcal A-\frac{2}{3}\mathcal B^{3/2}} }
{\sqrt{(\det \tilde{\mathrm{S}}_{\mathbf{u'' u''}})_{\mathbf u''_1}}} +
\frac{i\hbar~e^{\mathcal A+\frac{2}{3}\mathcal B^{3/2}} }
{\sqrt{(\det \tilde{\mathrm{S}}_{\mathbf{u'' u''}})_{\mathbf u''_2}}}=
-\mathrm{K_{SC}^{(2)}}(\mathbf{v}'', \mathbf{u}',T).
\end{array}
\end{equation}

Finally we consider the uniform formula evaluated exactly at the
caustic. To do so we rewrite Eq.~(\ref{equn}) using the uniform
Jacobian given by Eq.~(\ref{o3}):
\begin{equation}
\mathrm{K_{SC}^{UN}} (\mathbf{v}'', \mathbf{u}',T)=\frac{1}{2\pi}
\left[{\left(\frac{i\pi\hbar}{\lambda_+}\right)^{1/2}}
{\left(\frac{-i\hbar}{3{G'}}\right)^{1/3}} \right] ~e^{\mathcal A}
\int e^{y^3/3}dy. \label{equno3}
\end{equation}
Since $\frac{(-i)^{1/3}}{2\pi}\int e^{y^3/3}dy=e^{-2\pi
i/3}~\mathrm{f_i}(0)$, we find the same result as found previously
with the formulas of the Sects.~\ref{regular} and \ref{transitional}
calculated at phase space caustics.

%%%%%%%%%%%%%%%%%%%%%%%%%%%%%%%%%%%%%%%%%%%%%%%%%%%%%%%%%%%%%%%%%%%%%%
%%%%%%%%%%%%%%%%%%%%%%%%%%%%%%%%%%%%%%%%%%%%%%%%%%%%%%%%%%%%%%%%%%%%%%
\section{Final Remarks}
\label{fr}

Semiclassical approximations for the evolution operator seem to be
plagued by focal points and caustics in any representation. A
relatively simple way to derive improved expressions that avoid the
singularities of such quadratic approximations is provided by the
Maslov method. The method explores the fact that, for example, the
coordinate representation of the propagator, $\langle
x|K(T)|x'\rangle$ can be written as the Fourier transform of the
propagator in its dual representation, $\langle x|K(T)|x'\rangle =
\int \langle x|p\rangle \langle p|K(T)|x'\rangle dp$. If the
trajectory from $x'$ to $x$ in the time $T$ corresponds to a focal
point, we can still use this integral expression and the usual
quadratic approximation for $\langle p|K(T)|x'\rangle$, as long as
we perform the integral over $p$ expanding the exponents to third
order around the stationary point. This results in a well behaved
approximation for the coordinate propagator in terms of an Airy
function. In this paper we have shown that a similar procedure can
be applied to the coherent state representation and derived three
similar third order formulas that can be used depending on how far
the stationary trajectory is from the phase space caustics. Although
we have considered only systems with two degrees of freedom the
extension to higher dimensions is immediate. We note that a uniform
formula for the coherent state propagator was previously derived in
\cite{fricke} for a particular Hamiltonian.

The regular formula~(\ref{regform}) is the simplest of our three
approximations and consists of a sum over the same complex
trajectories that enter in the quadratic approximation. The
contribution of each trajectory is regularized by a term that avoids
divergences at phase space caustics. We emphasize that this
regularization deals just with the problem of caustics, so that we
still need to identify contributing and non-contributing
trajectories in order to get acceptable results. This approximation
holds as far as the contributing trajectories are not too close to
the caustics, otherwise the vicinities of different trajectories can
start to overlap and their contributions would be miscounted. The
transitional formula~(\ref{eqkf}) works exactly in this situation.
It involves the contribution of the PSC trajectory alone, and
therefore is valid only very close to the caustics. Finally, the
uniform formula~(\ref{uform}) is valid everywhere, near of far a
caustic. The formula we derived deals with the simplest topology of
caustics \cite{berryupstill}.

All three semiclassical formulas derived here involve the
calculation of third order derivatives of the action. We presented
an algorithm to evaluate these derivatives numerically in
Appendix~\ref{ap1}. Numerical results using these expressions will
presented in a future publication.

%%%%%%%%%%%%%%%%%%%%%%%%%%%%%%%%%%%%%%%%%%%%%%%%%%%%%%%%%%%%%%%%%%%%%%
%%%%%%%%%%%%%%%%%%%%%%%%%%%%%%%%%%%%%%%%%%%%%%%%%%%%%%%%%%%%%%%%%%%%%%
\appendix
\section{Derivatives of the Action $\tilde{\mathcal S}$}
\label{ap1}

In this appendix, we show how second and third derivatives of
$\tilde{\mathcal S}(\mathbf u', \mathbf u'', T)$ can be calculated
for a given trajectory. This procedure can be used with any set of
variables (for example, $(\mathbf u', \mathbf v'', T)$ or $(\mathbf
q', \mathbf q'', T)$) with minor modifications.

\subsection{The Tangent Matrix and the Tangent Tensor}

The equations of motion in the ${\bf u}$ and ${\bf v}$ variables can
be written in compact form as
\begin{equation}
\dot{r}_i = J_{ij} H'_j
\label{eqmovap2}
\end{equation}
where the vector ${\bf r}$ and the matrix $J$ are given by
\begin{eqnarray}
{\mathbf r} = \left(
      \begin{array}{c} u_x\\ u_y\\ v_x\\ v_y\\  \end{array}
    \right)
    \quad \mathrm{and} \quad
J =  \left(
      \begin{array}{cccc}
         0 & 0 & -i/\hbar & 0 \\
         0 & 0 & 0 & -i/\hbar \\
         i/\hbar & 0 & 0 & 0 \\
         0 & i/\hbar & 0 & 0 \\
      \end{array}
    \right),
\label{eqmovdef}
\end{eqnarray}
and
\begin{eqnarray}
    \tilde{H}^{'}_{i} =
    \frac{\partial \tilde{H}}{\partial r_i} \;.
\end{eqnarray}
Expanding Eq.~(\ref{eqmovap2}) up to second order around a reference
trajectory ${\bar{\mathbf r}}(t)$, we obtain
\begin{eqnarray}
\delta \dot{r}_i = J_{ij}~ H^{''}_{jk} ~\delta r_k + \frac{1}{2}
J_{ij} ~\delta r_l ~H^{'''}_{lkj} ~\delta r_k \;,
\label{eqmovexp}
\end{eqnarray}
where
\begin{eqnarray}
\left.
    \tilde{H}^{''}_{ij} =
    \frac{\partial^2 \tilde{H}}{\partial r_i  \partial r_j}
\right|_{\bar{\mathrm{\mathbf r}}} \qquad  \mathrm{and} \quad
\left.
    \tilde{H}^{'''}_{ijk}=
    \frac{\partial^3 \tilde{H}}{\partial r_i \partial r_j  \partial r_k}
\right|_{\bar{\mathrm{\mathbf r}}} \;.
\label{h2eh3}
\end{eqnarray}

The solution of Eq.~(\ref{eqmovexp}) can be expressed in terms of
the initial displacement $\delta \mathbf{r}(0)$ as
\begin{equation}
\delta r_i(t) = M_{ij}(t) ~\delta r_j(0) + \delta r_k(0)~
U_{kli}(t)~ \delta r_l(0)\;,
\label{eqtent}
\end{equation}
where the tangent matrix $M$ and the tangent tensor $U$ satisfy
$M(0)=\mathbf 1$ and $U(0)=0$. Differentiating this equation with
respect to $t$ and by using Eq.~(\ref{eqmovexp}), we obtain the
differential equations satisfied by $M$ and $U$ directly:
\begin{equation}
\begin{array}{l}
\dot{M}_{ij}(t) \delta r_j(0) + \delta r_k(0) \dot{U}_{kli}(t)
\delta r_l(0) = J_{ij}~ H^{''}_{jk} ~M_{kl} ~\delta r_l(0) ~ + \\
J_{ij}~ H^{''}_{jm}~\delta r_k(0) U_{klm}~\delta r_l(0) +
 \frac{1}{2} J_{ij} ~\delta r_k(0) ~M_{nk} ~H^{'''}_{nmj}~M_{ml} ~\delta
 r_l(0) \;,
\label{eqtentdot}
\end{array}
\end{equation}
where we have discarded terms of third order in $\delta r_i(0)$.
This leads to
\begin{equation}
\dot{M}_{ij} = J_{il}~ H^{''}_{lk} ~M_{kj} \label{eqm}
\end{equation}
and
\begin{equation}
\dot{U}_{ijk} = J_{kl}~ H^{''}_{lm} ~U_{ijm} + \frac{1}{2} J_{kl}
~M_{ni} ~H^{'''}_{nml}~M_{mj}
\label{equ} \;.
\end{equation}

These two sets of differential equations can be solved for a given
reference trajectory ${\bar{\mathbf r}}(t)$ and boundary
conditions $M (0) = \mathbf 1$ and $U (0)=0$.

\subsection{Derivatives of $\tilde{\mathcal S}$}

Here we show how to obtain the second and third derivatives of
$\tilde{\mathcal{S}}$ in terms of $M$ and $U$. We start from
Eqs.~(\ref{stilpartial}), which can be written as
\begin{equation}
V_i = K_{ij} \tilde{S}^{'}_j \label{b1} \;,
\end{equation}
where
\begin{eqnarray}
{\mathbf V} =
   \left(
     \begin{array}{c}
       {v}'_{x}\\ {v}'_{y}\\ {v}''_{x}\\ {v}''_{y}\\
     \end{array}
   \right), \quad
{\mathbf U} =
   \left(
     \begin{array}{c}
       {u}'_{x}\\ {u}'_{y}\\ {u}''_{x}\\ {u}''_{y}\\
     \end{array}
   \right), \quad
K =
   \left(
     \begin{array}{cccc}
       i/\hbar & 0 & 0 & 0 \\
       0 & i/\hbar & 0 & 0 \\
       0 & 0 & -i/\hbar & 0 \\
       0 & 0 & 0 & -i/\hbar \\
     \end{array}
   \right)
\label{b2}
\end{eqnarray}
and $\tilde{S}^{'}_i = \partial \tilde{\mathcal S}/\partial U_i$.

Considering variations on Eq.~(\ref{b1}) around the reference
trajectory and expanding up to second order, we get
\begin{eqnarray}
\delta V_i = K_{ij}~ \tilde{S}^{''}_{jk} ~\delta U_k + \frac{1}{2}
K_{ij} ~\delta U_l ~\tilde{S}^{'''}_{lkj} ~\delta U_k \;,\label{b3}
\end{eqnarray}
where
\begin{eqnarray}
\left.
    \tilde{S}^{''}_{ij} =
    \frac{\partial^2 \tilde{\mathcal S}}{\partial U_i  \partial U_j}
\right|_{\bar{\mathrm{\mathbf r}}} \qquad  \mathrm{and} \quad
\left.
    \tilde{S}^{'''}_{ijk}=
    \frac{\partial^3 \tilde{\mathcal S}}{\partial U_i \partial U_j  \partial U_k}
\right|_{\bar{\mathrm{\mathbf r}}} \;.
\label{b4}
\end{eqnarray}
The idea now is to manipulate Eq.~(\ref{b3}) so that
final displacements are written in terms of the initial ones. To
do this we write
\begin{equation}
\begin{array}{ll}
\delta {\mathbf U} &= A \delta {\mathbf r}(0) + B \delta {\mathbf r}(T) \\
\delta {\mathbf V} &= C \delta {\mathbf r}(0) + D \delta {\mathbf
r}(T)
\end{array}
\label{b5}
\end{equation}
where $A$, $B$, $C$ and $D$ are $4\times 4$ matrices that can be
written in terms of $2\times 2$ blocks as
\begin{eqnarray}
A = \left( \begin{array}{cc} {\mathbf 1} & {\mathbf 0} \\ {\mathbf
0} & {\mathbf 0}
\end{array} \right), \qquad B = \left( \begin{array}{cc} {\mathbf 0} & {\mathbf
0} \\
{\mathbf 1} & {\mathbf 0}
\end{array} \right), \qquad C = \left(\begin{array}{cc} {\mathbf 0} & {\mathbf 1}
\\ {\mathbf 0}
& {\mathbf 0}
\end{array} \right), \qquad D = \left( \begin{array}{cc} {\mathbf 0} & {\mathbf
0} \\
{\mathbf 0} & {\mathbf 1} \end{array} \right). \label{b6}
\end{eqnarray}
Replacing Eqs.~(\ref{b5}) into (\ref{b3}) and solving for $\delta
{\bf r}(T)$ produces
\begin{eqnarray}
\begin{array}{l}
      \delta \mathbf{r}(T) =
      \left(
        D  - K \, \tilde{\mathrm{S}}'' \, B
      \right)^{-1}
      \left(
        K \, \tilde{S}'' \, A  - C
      \right)
      \delta \mathbf{r}(0)  + \frac{1}{2}
      \Lambda^{-1}
       \, \mathbf{w},
\end{array}
\label{b7}
\end{eqnarray}
where $\Lambda \equiv K^{-1}(D-K \tilde{S}''B)$ and
\begin{equation}
\begin{array}{ll}
w_i &= \delta u_l \tilde{S}^{'''}_{lmi} \delta u_m \\
 &= [A \delta {\mathbf r}(0) + B \delta {\mathbf r}(T)]_l~ \tilde{S}^{'''}_{lmi}
~
    [A \delta {\mathbf r}(0) + B \delta {\mathbf r}(T)]_m \\
 &\approx [A \delta {\mathbf r}(0) + B M \delta {\mathbf r}(0)]_l~
\tilde{S}^{'''}_{lmi} ~
    [A \delta {\mathbf r}(0) + B M \delta {\mathbf r}(0)]_m \\
 &= [L \delta {\mathbf r}(0)]_l~ \tilde{S}^{'''}_{lmi} ~ [L \delta {\mathbf
r}(0)]_m
 \;.
 \label{b8}
\end{array}
\end{equation}
In this expression we have discarded terms of third order in $\delta
{\mathbf r}(0)$ and we have defined the auxiliary matrix $L = A+BM$.
Computing all these matrices explicitly, we find
\begin{eqnarray}
\left( D  - K \, \tilde{S}''  \, B  \right)^{-1} =
   \left(
      \begin{array}{cc}
        i \hbar \tilde{S}^{-1}_{\mathbf{u}'  \mathbf{u}''} & 0 \\
        \tilde{S}_{\mathbf{u}'' \mathbf{u}''}\,
          \tilde{S}^{-1}_{\mathbf{u}' \mathbf{u}''} & 1
      \end{array}
   \right), \\
\left( K  \, \tilde{S}'' \, A -C \right) =
   \left(
      \begin{array}{cc}
        (i/\hbar)\tilde{S}_{\mathbf{u}'  \mathbf{u}'} & -1 \\
        - (i/\hbar)\tilde{S}_{\mathbf{u}''  \mathbf{u}'} & 0
      \end{array}
   \right),
   \label{b9}
\end{eqnarray}
\begin{eqnarray}
L^{-1} =
    \left(
      \begin{array}{cc}
        1 & 0 \\
        -M^{-1}_{\mathbf{uv}} M_{\mathbf{uu}} & M^{-1}_{\mathbf{uv}}
      \end{array}
   \right), \quad
\Lambda = -i \hbar
   \left(
      \begin{array}{cc}
         -M_{\mathbf{uv}}^{-1} & 0 \\
         M_{\mathbf{vv}}~ M_{\mathbf{uv}}^{-1}& -1
      \end{array}
   \right)\;.
   \label{b10}
\end{eqnarray}

Comparing linear terms of Eq.~(\ref{b7}) with (\ref{eqtent}), we
find
\begin{equation}
M = \left(
\begin{array}{cc}
  M_{\mathbf{uu}} & M_{\mathbf{uv}} \\
  M_{\mathbf{vu}} & M_{\mathbf{vv}} \end{array}
\right) = \left(
  \begin{array}{cc}
    - \tilde{\mathrm{S}}^{-1}_{\mathbf{u}'  \mathbf{u}''} \,
       \tilde{\mathrm{S}}_{\mathbf{u}'  \mathbf{u}'} &
       -i \hbar \tilde{\mathrm{S}}^{-1}_{\mathbf{u}'  \mathbf{u}''} \\
    (i/\hbar)
       \left(
          \tilde{S}_{\mathbf{u}''  \mathbf{u}''} \,
          \tilde{S}^{-1}_{\mathbf{u}'  \mathbf{u}''} \,
          \tilde{S}_{\mathbf{u}'  \mathbf{u}'} -
           \tilde{S}_{\mathbf{u}''  \mathbf{u}'} \right) \quad &
       - \tilde{S}_{\mathbf{u}''  \mathbf{u}''} \,
         \tilde{S}^{-1}_{\mathbf{u}'  \mathbf{u}''}
  \end{array}
\right)
\label{b11}
\end{equation}
or
\begin{eqnarray}
\tilde{ S}'' = \left(
    \begin{array}{cc}
       \tilde{S}_{\mathbf{u}'  \mathbf{u}'} &
         \tilde{S}_{\mathbf{u}'  \mathbf{u}''} \\
       \tilde{S}_{\mathbf{u}''  \mathbf{u}'} &
         \tilde{S}_{\mathbf{u}''  \mathbf{u}''}
    \end{array}
\right) =
     i\hbar \left(
   \begin{array}{cc}
     M_{\mathbf{u} \, \mathbf{v}}^{-1} \,
       M_{\mathbf{u} \, \mathbf{u}}&
       - M_{\mathbf{u}\,\mathbf{v}}^{-1} \\
       -
        \left(
         M_{\mathbf{v} \,\mathbf{v}}  \,
           M_{\mathbf{u} \, \mathbf{v}}^{-1} \,
         M_{\mathbf{u} \, \mathbf{u}} + M_{\mathbf{v} \, \mathbf{u}}
       \right) \quad &
      M_{\mathbf{v} \,\mathbf{v}} \,
       M_{\mathbf{u} \, \mathbf{v}}^{-1}
   \end{array}
\right). \label{b12}
\end{eqnarray}

Finally, comparing the quadratic terms,
\begin{equation}
\begin{array}{ll}
\frac{1}{2} \Lambda^{-1}_{ik} w_k &= \frac{1}{2}
\Lambda^{-1}_{ij}~L_{nk} ~\delta r_k ~\tilde{S}^{'''}_{nmj}~ L_{ml}~ \delta r_l
\\
&\equiv \delta r_k ~U_{kli}~ \delta r_l
\end{array}
\end{equation}
or
\begin{equation}
\frac{1}{2} \Lambda^{-1}_{ij}~L_{nk} ~\tilde{S}^{'''}_{nmj}~
L_{ml} = U_{kli} \;.
\end{equation}
Solving for the third derivatives of $\tilde{S}$ produces
\begin{equation}
\tilde{S}^{'''}_{ijk} = 2 L^{-1}_{mi}~
\Lambda_{kn}~U_{mln}~L^{-1}_{lj},
\end{equation}
where $\Lambda$ and $L^{-1}$ are given by (\ref{b10}).

%%%%%%%%%%%%%%%%%%%%%%%%%%%%%%%%%%%%%%%%%%%%%%%%%%%%%%%%%%%%
\section*{Acknowledgments}

MAMA and ADR acknowledge financial support from CNPq, FAPESP and
FINEP. ADR especially acknowledges FAPESP for the fellowship $\#$
00/00063-2 and 04/04614-4, and also A.F.R. de Toledo Piza for
stimulating discussions.

%%%%%%%%%%%%%%%%%%%%%%%%%%%%%%%%%%%%%%%%%%%%%%%%%%%%%%%%%%%%
%%%%%%%%%%%%%%%%%%%%%%%%%%%%%%%%%%%%%%%%%%%%%%%%%%%%%%%%%%%%

\end{document}